# OPTICAL ABSORPTION AND EMISSION OF A QUANTUM DOT IN THE KONDO REGIME

PARTHA GOSWAMI
*Deshbandhu college, University of Delhi, Kalkaji, New Delhi-110019, India*

GARIMA PUNEYANI
*Department of Electronic Sciences, University of Delhi South Campus, Delhi-110021,India*

AVINASHI KAPOOR
*Department of Electronic Sciences, University of Delhi South Campus,Delhi-110021,India*

**Abstract.** A variant of the Anderson model, that describes hybridization between localized state (c-state) of a quantum dot and a Fermi sea conduction band, is investigated. We demonstrate that, as a function of the hybridization parameter v, the system undergoes a crossover from the state where the conduction band and the c-level are fully coupled to a state where these are decoupled. The c-electron spectrum, however, has a gap together with the presence of the Kondo peak in the former state. For the latter, we have a Mott-like localization where the c-electron spectrum again has a gap without the Kondo peak. Within this gap the conduction electrons fully recover the free band density of states and the effective hybridization is practically zero. Our main aim, however, is to study the emission and absorption in a quantum dot with strongly correlated Kondo ground state. We use the Green's function equation of motion method for this purpose. We calculate the absorption/emission (A/E) spectrum in the Kondo regime through a symmetrized quantum autocorrelation function obtainable directly within perturbation theory using the Fermi golden rule approximation. The spectrum reveals a sharp, tall peak close to Kondo-Abrikosov-Suhl peak and a few smaller, distinguishable ones on either side. The former clearly indicates that the Kondo phenomenon has its impact on A/E ( non-Kondo processes), which are driven by the coupling involving the dipole moment of quantum dot transitions reflecting the physical structure of the dot including the confinement potential, in the Kondo regime.

## I. INTRODUCTION

It is feasible to measure the absorption and emission and absorption spectrum of quantum dots (QDs) ( for example, self-assembled InAs dots embedded in GaAs) in a number of optical experiments[1,2].In the emission spectrum measurements, an exciton created by illumination recombines inside the QD, whereby a photon is emitted which is measured. In absorption spectrum measurements, on the other hand, photons are absorbed inside the QD by electron-hole pair (exciton) excitation. Due to quantum structural confinement, the QD possesses a discrete energy level structure, which can be rigidly shifted with respect to the Fermi energy $E_F$ by tuning, say, dipole moment of QD transition. The optical data [1,2] justify the assumption of a discrete energy level structure of the QD. Motivated by these experimental findings we investigated a variant of Anderson model [3] which describes the non-local Fermi sea conduction band, lowest local conduction

state(c-state) within the QD and the highest local valence state(v-state). The reasons for considering only two levels are observations of phenomena related to the interaction of photons with discrete states in self-assembled dots, such as ground state Rabi oscillations[4], weak[5,6,7] and strong[8,9] coupling regimes in various micro-cavity structures, which have strengthened the picture of the optical electron-hole pair excitation in a QD as a coherent two-level transition. Here, we focus on the optical absorption/emission, which are non-Kondo processes, in the Kondo regime. We are interested, in particular, on how Kondo correlations resulting from the presence of the Fermi sea affect the absorption and emission spectra. The optical absorption of electromagnetic radiation by this two-level transition is mediated via a coherent superposition of the initial and final state, the polarization. We calculate a symmetrized quantum autocorrelation function obtainable within perturbation theory using the Fermi golden rule approximation. This function gives the frequency dependent transition rate corresponding to absorption and emission at a finite temperature. We find that absorption and emission spectra of a QD correspond to a strongly correlated Kondo ground state. The numerical renormalization group method, possibly is the best available one to calculate the absorption (emission) spectrum of a QD which starts from (ends up in) a strongly correlated Kondo ground state. We have, however, deliberately adopted much simpler and reliable equation of motion (EOM) method. The physical motive is to formulate a simple and sound theoretical basis of optical absorption/emission in a confined system with the inclusion of non-locality and to give a reliable prediction of the line-shift and peak reduction. We predict that the absorption/emission spectra consist of delta peaks only.

The Kondo effect is the coherent coupling of a single unpaired electron spin with a Fermi sea of electrons around this single spin. The key feature of the Kondo effect is the formation of the Kondo-Abrikosov-Suhl(KAS) resonance near the Fermi level. The scanning tunneling microscopy (STM) technique allowed visualization of this resonance for magnetic impurities. The resonance is a consequence of the strong quantum correlations between the localized magnetic moment and the conduction electrons. The spins of the surrounding electrons screen this single spin effectively forming a singlet. In a nut-shell, an odd, unpaired electron in a strongly coupled quantum dot makes the dot to behave as a magnetic impurity screened by delocalized electrons. Such a Kondo impurity creates a KAS peak in the local density of states (LDOS) at the Fermi level, thereby leading to characteristic Kondo resonances with enhanced conductance around zero bias, which has been observed in various quantum dot systems during the recent years [10,11,12,13]. We too show in this paper the occurrence of KAS peak close to Fermi level. At a given temperature, as the hybridization parameter v tends towards $0^+$, the c-fermion DOS (or LDOS) show both a suppression of the peak close to Fermi level and a shift to less negative energies. We demonstrate that, as a function of the tuning parameter v, the system undergoes a crossover from the state where the conduction band and the c-level are fully coupled to a state where these are decoupled. To clarify, for $v > v_c \rightarrow 0^+$ ($v_c \approx 0.01$) the system is in a phase where the c- electrons present a Kondo peak with infinite lifetime and reasonable peak height close to the Fermi level and take active part in the conduction. The c-electron spectrum, however, has a gap $\Delta$ together with the presence of the Kondo peak. For $v < v_c$ we have a Mott-like localization, where the c-electron spectrum again has a gap $\Delta$ without the Kondo peak. Within this gap the conduction electrons fully recover the free band DOS and the effective hybridization is practically zero. This shows that the conduction band at low energy is completely decoupled from the c-level, but the effective hybridization remains active at a finite intermediate energy scale. It is also noticed that starting from $v_c$ as the tuning parameter v is increased there is an overall decrease in the Mott gap $\Delta$. Furthermore, the increase in temperature leads to the peak reduction in the KAS resonance(the peak-broadening does not occur as this is not possible in our EOM framework). The effect is reminiscent of the de-coherence caused by the measurement leading to the pure decrease of the amplitude of the resonance without any smearing. The destruction of the Kondo effect by a decohering action of an external microwave field has been considered in Ref.[14]. Here we shall show that the external radiation, which works similar to changing temperature, leads to dissipation-less reduction (or decoherence) of the Kondo peak. The absence of dissipation ( broadening/smearing of the Kondo peak) is, in fact, a consequence of the Hartee-Fock approximation framework adopted here.  It may be mentioned that the finite dc voltage is also a source of de-coherence. Our future aim is to  express linear conductance and equilibrium current fluctuations in terms of the single-particle Green function(see Ref.15) This relation  is believed to become exact at low frequencies, when charge fluctuations on the dot can be neglected. It has been shown to work very well for frequencies of the order of the Kondo scale. This opens up the possibility of measuring the equilibrium Kondo resonance directly in a transport measurement

The paper is organized as follows: In Section 2 we present a variant of Anderson model which describes the hybridization between Fermi sea conduction band with a local conduction level(c-level) of quantum dot and includes a local valence band level as well. With this model we investigate the transition of the system from a phase where the c-electrons present a Kondo peak with infinite lifetime and reasonable peak height close to the Fermi level to a phase characterized by a Mott-like localization where the c-electron spectrum again has a gap $\Delta$ without the Kondo peak. We also find that the involvement of the dipole moment of the QD transition enhances $\Delta$ in comparison with the case when the involvement is absent. The Anderson Hamiltonian is though solvable through the Bethe Ansatz , Numerical Renormalization Group  and Quantum Monte Carlo methods  but a reliable and simple method to obtain dynamical properties at low temperatures  is not available.  A symmetrized  quantum autocorrelation function which yields  the frequency dependent transition rate corresponding to absorption and emission will be calculated in Section 3 by equation of motion method  to address this need. The  paper ends with a discussion  on the possible use of the investigation carried out in quantum dot solar cell(QDSC) in  Section 4.

## II. ANDERSON MODEL AND SINGLE PARTICLE SPECTRUM

The model Hamiltonian H  under consideration consists of four terms:

$$H = H_{FS} + H_{hybrid} + H_{c-v} + H_{pert} . \tag{1}$$

$$H_{FS} = \Sigma_{k,\sigma}(\varepsilon_k - \sigma h)\, d^{\dagger}_{k\sigma}\, d_{k\sigma} \,,\quad H_{hybrid} = \Sigma_{k,\sigma} V_k\, (d^{\dagger}_{k\sigma} c_{\sigma} + c^{\dagger}_{\sigma} d_{k\sigma}) \tag{2}$$

$$H_{c\text{-}v} = \Sigma_\sigma \varepsilon_c\, n_{c,\sigma} + \Sigma_\sigma \varepsilon_v\, n_{v\sigma} + U_c\, n_{c\uparrow} n_{c\downarrow}$$

$$+ U_v\, (1 - n_{v\uparrow})(1 - n_{v\downarrow}) - U_{coul} \Sigma_{\sigma, \sigma'}\, n_{c,\sigma}(1 - n_{v,\sigma'}). \tag{3}$$

The term ($\varepsilon_k - \sigma h$) with $h \to 0^+$ corresponds to a ferromagnetic Fermi sea(FS) (electrode) to support the spin polarization. The Fermi operator $d^\dagger_{k,\sigma}$ creates a delocalized spin $\sigma$ electron with wave vector $\mathbf{k}$. The hybridization between the c-level and FS is described by $H_{hybrid}$. In what follows we shall assume the hybridization parameter $V_k$ to be real and k- independent. Moreover, since in most experiments essentially only a level close to the Fermi energy are relevant, we shall take later a single vector $\mathbf{k}$ ( $|\mathbf{k}| \approx |\mathbf{k_F}|$, with $\mathbf{k_F}$ the Fermi momentum) as a minimal model to compensate a localized dot(with c- and v- levels) spin which corresponds to an impurity site here. The model thus resembles a single impurity Anderson problem.

The quantities $\varepsilon_c$ and $\varepsilon_v$, respectively, correspond to c-level and v-level energies. It may be noted that $\varepsilon_v$ is smaller than $\varepsilon_c$ by the order of the bulk band gap $\varepsilon_G$. Their precise values are not important, except for setting the overall scale for the threshold for absorption or emission processes. Here $n_{c\sigma} \equiv c^\dagger_\sigma c_\sigma$ and $n_{v\sigma} \equiv v^\dagger_\sigma v_\sigma$. The Fermi operators $c^\dagger_\sigma$ and $v^\dagger_\sigma$ create a spin-$\sigma$ electron in a c-level or in a v-level, respectively. To include the coulomb interactions we have introduced the parameters $U_c$ and $U_v$. These are large Coulomb repulsion energies which have to be paid if a c-level is occupied by two electrons or if a v-level is empty, respectively. The term $[- U_{coul} \Sigma_{\sigma, \sigma'}\, n_{c,\sigma}(1 - n_{v,\sigma'})]$ accounts for the exciton binding. The last part of the Hamiltonian $H_p$ corresponds to the perturbation as explained below:

$$H_{pert} = \Sigma_{k,\sigma} \{\mu_k\, a_k \exp(-i\omega_k t)\, c^\dagger_\sigma v_\sigma + \mu_k\, a^\dagger_k \exp(i\omega_k t)\, v^\dagger_\sigma c_\sigma \}. \tag{4}$$

This describes the excitation (first term) and the annihilation (second term) of excitons in the QD by photon absorption or photon emission, respectively. Here $a_k$ ($a^\dagger_k$) destroys (creates) a photon of the illuminating field with wave vector $\mathbf{k}$, where the photon has the frequency $\omega_k = c|\mathbf{k}|$. The coupling is given by $\mu_k = e(\hbar \omega_k / 2\varepsilon_0 \Omega)^{1/2} \mathbf{e_k} \cdot \mathbf{D}$, with the elementary charge e, the dielectric constant $\varepsilon_0$, the quantization volume $\Omega$, the orientation of the photon field $\mathbf{e_k}$, and the dipole moment $\mathbf{D}$ of the QD transition. We assume $\mu$ to be independent of $\mathbf{k}$. The theoretical calculation [16] and the experimental investigation[17]of the polarization-dependent well-known Rabi oscillations [18, 19, 20, 21] using photoluminescence spectroscopy have shown that the dipole moment directly reflects the physical structure of the dot including the confinement potential. In fact, in the time domain, the strong field interaction leads to the well-known Rabi oscillations. In the frequency domain, it will introduce Rabi side bands in the absorption. This phenomenon has been studied theoretically [22,23,24] very extensively in early seventies and demonstrated experimentally in atomic systems [25,26] in late seventies and eighties. Recently, these effects have also been observed in molecular systems and quantum dot [27, 28, 29, 30]. In this communication we shall not take up this issue. We shall also not take up the issue of scattering of either electron or hole by phonons (see also Section 4).Furthermore, the quantized nature of the photon field will be assumed not to play any role ; to calculate emission and absorption line shapes, all that we shall be concerned with are the Green's functions corresponding to the operators ($c^\dagger_\sigma v_\sigma + v^\dagger_\sigma c_\sigma$). For simplicity of notation, we shall therefore henceforth write the term in (4) simply as $H_{pert} = \mu \sum_\sigma (c^\dagger_\sigma v_\sigma + v^\dagger_\sigma c_\sigma)$. We have neither introduced a Zeeman term nor an interaction term involving cyclotron frequency of the semi-conductor and single-particle angular momentum. The reasons being these terms are non-zero only in the presence of a magnetic field.

The presence of atomic like spectra of quantum dots calls for appropriate theoretical tools of investigation. One may treat the electronic states of the dot using Density Functional Theory[31] with the explicit involvement of exchange interaction [32,33]. However, the spin-dependent properties are not the focal point of the investigation at present. Besides, the strong coulomb repulsions ($U_c, U_v$) are to be handled consistently. We, therefore, prefer the equation of motion (EOM) method (based on Hartree-Fock approximation (HFA)) which is known to be reliable in the coulomb blockade regime and qualitatively correct for Kondo Physics[34,35]. It must be noted that the exact diagonalization method [36,37] and HFA scheme [38,39] have their own limitations. For example, quasi-particle lifetime will be infinite in HFA as the HF Hamiltonian is quadratic in fields and therefore exactly diagonalizable. This means the corresponding eigenstates are stationary states with infinite lifetime. As a scoring point, however, it may be noted that solutions corresponding to HFA scheme become exact in the U→0 limit. In the opposite limit U→∞, double occupation of the c- and v-levels are forbidden, and fermion number cannot fluctuate by two. Upon considering only the coulomb interactions induced pairing of opposite spins, the Hamiltonian in (1) can be written as

$$H_1 \approx H_0 + H_I \tag{5}$$

$$H_0 = U_v + \sum_k (\varepsilon_k - h) d^\dagger_{k\uparrow} d_{k\uparrow} + \sum_k (\varepsilon_k + h) d^\dagger_{k\downarrow} d_{k\downarrow} + (\varepsilon_c - U_{coul}) c^\dagger_\downarrow c_\downarrow$$

$$+ (\varepsilon_c - U_{coul} + U_c) c^\dagger_\uparrow c_\uparrow + \varepsilon_v v^\dagger_\uparrow v_\uparrow + (\varepsilon_v - U_v) v^\dagger_\downarrow v_\downarrow$$

$$H_I = \sum_k V_k (d^\dagger_{k\uparrow} c_\uparrow + c^\dagger_\uparrow d_{k\uparrow} + d^\dagger_{k\downarrow} c_\downarrow + c^\dagger_\downarrow d_{k\downarrow})$$

$$+ \sum_k V_k (d^\dagger_{k\uparrow} v_\uparrow + v^\dagger_\uparrow d_{k\uparrow} + d^\dagger_{k\downarrow} v_\downarrow + v^\dagger_\downarrow d_{k\downarrow})$$

$$+ \mu (c^\dagger_\uparrow v_\uparrow + v^\dagger_\uparrow c_\uparrow + c^\dagger_\downarrow v_\downarrow + v^\dagger_\downarrow c_\downarrow) + \Delta_c c^\dagger_\downarrow c_\uparrow + (\Delta^\dagger_c + U_c) c^\dagger_\uparrow c_\downarrow$$

$$+ \Delta_v v^\dagger_\downarrow v_\uparrow + (\Delta^\dagger_v + U_v) v^\dagger_\uparrow v_\downarrow + (\Delta_1 v^\dagger_\downarrow c_\uparrow + \Delta_2 v^\dagger_\uparrow c_\downarrow + h.c.).$$

which is completely diagonal in FS, c-and v-level operators. Here

$$\Delta_c = -U_c \langle c^\dagger_\uparrow c_\downarrow \rangle, \Delta^\dagger_c = -U_c \langle c^\dagger_\downarrow c_\uparrow \rangle, \Delta_v = -U_v \langle v^\dagger_\uparrow v_\downarrow \rangle,$$

$$\Delta^\dagger_v = -U_v \langle v^\dagger_\downarrow v_\uparrow \rangle, \Delta_1 = -U_{coul} \langle c^\dagger_\uparrow v_\downarrow \rangle, \Delta_2 = -U_{coul} \langle c^\dagger_\downarrow v_\uparrow \rangle . \tag{6}$$

These averages will now be calculated in a self-consistent manner below. Note that we have made static-path-approximation (SPA) for quantum auxiliary fields $\Delta_c = = -U_c \langle c^\dagger_\uparrow c_\downarrow \rangle$, $\Delta_v = -U_v \langle v^\dagger_\uparrow v_\downarrow \rangle$, etc..The SPA describes exactly classical fluctuations. In order to account for quantum fluctuations one should take Δ's time-dependent. We proceed with finite-temperature formalism here. Since the Hamiltonian $H_1$ is completely diagonal one can write down easily the equations for the operators $\{d_{k\sigma}(\tau), c_\sigma(\tau), v_\sigma(\tau),...\}$, where the time evolution an operator O is given by $O(\tau) = \exp(H\tau) O \exp(-H\tau)$, to ensure that the thermal averages in $H_1$ are determined in a self-consistent manner. The Green's functions $G_c(\sigma,\tau) = -\langle T\{c_\sigma(\tau) c^\dagger_\sigma(0)\}\rangle$ and $G_v(\sigma,\tau) = -\langle T\{v_\sigma(\tau) v^\dagger_\sigma(0)\}\rangle$, where T is the time-ordering operator which arranges other operators from right to left in the ascending order of time $\tau$, are of primary interest as the poles of the Fourier transform of these functions yield the single-particle excitation spectra.

In the EOM approach, one can derive the equation for the temperature Green functions above by differentiating with respect to imaginary time $\tau$, with new Green functions generated in the equation. To close the equation chain, we have already diagonalized the Hamiltonian above in a meaningful manner. We neglect the hybridization between the v-level and FS assuming that the mass of the holes are significantly larger than the mass of the electrons. With this assumption, EOM for the c- and v- operators can be easily written down which give the EOM of the function $G_c(\uparrow,\tau)$:

$$(\partial/\partial\tau) G_c(\uparrow,\tau) = -(\varepsilon_c + U_c) G_c(\uparrow,\tau) - V_k G_{dc}(\uparrow,\tau) - \mu G_{vc}(\uparrow,\tau)$$

$$- (\Delta^\dagger_c + U_c) G_c(\downarrow\uparrow,\tau) - \Delta^\dagger_1 G_{exc}(\downarrow\uparrow,\tau) - \delta(\tau), \qquad (7)$$

where

$$G_{dc}(\uparrow,\tau) = -\langle T\{d_{k\uparrow}(\tau) c^\dagger_\uparrow(0)\}\rangle, \quad G_{vc}(\uparrow,\tau) = -\langle T\{v_\uparrow(\tau) c^\dagger_\uparrow(0)\}\rangle,$$

$$G_c(\downarrow\uparrow,\tau) = -\langle T\{c_\downarrow(\tau) c^\dagger_\uparrow(0)\}\rangle, \quad G_{exc}(\downarrow\uparrow,\tau) = -\langle T\{v_\downarrow(\tau) c^\dagger_\uparrow(0)\}\rangle. \qquad (8)$$

The equations of motion of the new temperature functions can be written down in similar manner. As it is clear from above, the number of equations to deal with is five. Thus, in order to get single-particle excitation spectrum we shall have to solve a quintic. The Fourier coefficients of the thermal averages above are the Matsubara propagators $\{G_c(\uparrow, z´), G_{dc}(\uparrow, z´), G_c(\downarrow\uparrow,z´), G_{exc}(\uparrow\downarrow, z´), G_{vc}(\uparrow, z´)\}$ where $i\omega_n = z´ = [(2n+1)\pi i/\beta]$ with $n = 0, \pm 1, \pm 2,...$. Assuming $V_k$ independent of k, it is tedious but straightforward to see from (7),(8),... that the Fourier coefficient $G_c(\uparrow, z´) = (D_1/D_0)$. Before we write explicit expressions for $D_1$ and $D_0$, we linearize the dispersion relation $\varepsilon_k$ of the conduction band. In fact, we wish to cast the expressions into forms that are more convenient for a numerical analysis. The linearization of the dispersion relation gives $\varepsilon_k = Dk$. The wave-number k runs from $-1$ to 1, therefore 2D is the width of the conduction band. This assumption is equivalent to adopting a constant density of states, $\acute{\Gamma} = 1/(2D)$. Since, we approximated the c-level and conduction band coupling with a k-independent hybridization strength V, the level width $\Gamma = \pi\acute{\Gamma} V^2$. Neither of these assumptions affects the results in a significant way. We have, by and large, followed the notations in refs. 40 and 41. We also assume wide band limit [see Fig.1 where the bands correspond to $\{(\varepsilon_c + U_c), \varepsilon_F, \varepsilon_c, \varepsilon_v, (\varepsilon_v - U_v)\}$ and the quantities $(\varepsilon_F, \varepsilon_c, \varepsilon_v, U_c, U_v, U_{coul})$ are $\varepsilon_c = -(D/6), \varepsilon_F = 0, \varepsilon_v = -(D/3), U_c = (D/3), U_v = (2D/3)$, and $U_{coul} = (D/6)$] where D is greater than all other energy scale in the problem. The hybridization parameter (V) is assumed to be (D/18) in Fig.1; the separation between ($\varepsilon_c$/D) and (($\varepsilon_c$+$U_c$)/D) is 0.3333. The quantity h is set equal to zero. With these assumptions we find for k =0 and z = (z´/D)

$$(D_0 / D^5) = z\,[(z + (1/3))(z +1)(z +(1/6))(z -(1/6))$$

$$+ (1/9)\,(z +(1/3))(z +1)\langle c^\dagger_\uparrow c_\downarrow\rangle(1 - \langle c^\dagger_\downarrow c_\uparrow\rangle)$$

$$+ (4/9)(z +(1/6))(z -(1/6))\,\langle v^\dagger_\uparrow v_\downarrow\rangle(1- \langle v^\dagger_\downarrow v_\uparrow\rangle)$$

$$+(4/81)\,\langle c^\dagger_\uparrow c_\downarrow\rangle(1 - \langle c^\dagger_\downarrow c_\uparrow\rangle)\,\langle v^\dagger_\uparrow v_\downarrow\rangle(1- \langle v^\dagger_\downarrow v_\uparrow\rangle)$$

$$- (1/108)\,(z +(1/3))\,(1 - \langle c^\dagger_\downarrow c_\uparrow\rangle)\langle c^\dagger_\downarrow v_\uparrow\rangle\,\langle v^\dagger_\uparrow c_\downarrow\rangle$$

$$- (1/36)\,(z +(1/3))\,(z +(1/6))\,\langle v^\dagger_\downarrow c_\uparrow\rangle\,\langle c^\dagger_\uparrow v_\downarrow\rangle$$

$$- (1/162)\,\langle c^\dagger_\uparrow v_\downarrow\rangle\,\langle c^\dagger_\downarrow v_\uparrow\rangle\,\langle c^\dagger_\uparrow c_\downarrow\rangle\,\langle v^\dagger_\uparrow v_\downarrow\rangle$$

$$+ (1/54)(z -(1/6))\,\langle v^\dagger_\uparrow v_\downarrow\rangle\langle c^\dagger_\downarrow v_\uparrow\rangle\,\langle v^\dagger_\downarrow c_\uparrow\rangle]$$

$$- (1/324)\,[(z + (1/3))\,(z +1)\,(z + (1/2) - (1/3)\langle c^\dagger_\downarrow c_\uparrow\rangle)$$

$$- (4/9)\,\langle v^\dagger_\uparrow v_\downarrow\rangle\,(1- \langle v^\dagger_\downarrow v_\uparrow\rangle)\,(z +(1/2) - (1/3)\langle c^\dagger_\downarrow c_\uparrow\rangle)$$

$$- (1/54)\langle v^\dagger_\uparrow v_\downarrow\rangle\,\langle c^\dagger_\downarrow v_\uparrow\rangle\,(\langle v^\dagger_\uparrow c_\downarrow\rangle - \langle v^\dagger_\downarrow c_\uparrow\rangle)\,]. \qquad (9)$$

Similarly, an expression for $D_1$ can be written down. The dipole moment **D** of the QD transition has been assumed to be vanishingly small. The thermodynamic averages in (9) are to be determined using Hamiltonian in (5).

Nearly a decade ago, the Kondo effect was observed in a lithographically defined quantum dot, in which the magnetic moment stems from a single unpaired spin. The discovery of the Kondo effect in artificial atoms spurred a revival in the study of Kondo physics [42, 43], due in part to the unprecedented control of relevant parameters in these systems. We shall now see that in the Kondo ( local moment) regime (9) reduces to a simple, workable form. For this regime, characterized by single occupancy, we have $-U_c + \Gamma \leq \varepsilon_c \leq \Gamma$ and ($U_c/\pi\Gamma$) $\gg$ 1. The quantity ($U_c/\pi\Gamma$) equals 21.8796 ( greater than one) in our case. Thus, we have $\langle c^\dagger_\uparrow c_\downarrow\rangle \approx 1 \approx \langle v^\dagger_\uparrow v_\downarrow\rangle$ provided that $-0.3285 \leq (\varepsilon_c/D) \leq 0.0048$. The hybridization parameter (V) is the only band alteration agent. In this case ($D_0/D^5$) in Eq.(9) corresponds to

$$(z + (1/3))(z +1)\,[\,z^3 - ((1/3) - 2a)\,z^2 + (a^2 - (a/3) - (1/324))z - ((a/324)+(1/972))] \approx 0$$

$$(10)$$

where $a = |(\varepsilon_c /D)|$. For $(\varepsilon_c /D) = (-1/6)$, we find that $(D_0 / D^5) = 0$ has five real roots given by $z_1 = (-1/3)$, $z_2 = (-1)$, $z_3 = 0.1411$, $z_4 = (-0.1967)$, and $z_5 = 0.0556$. The first two roots correspond to $(\varepsilon_v /D)$ and $((\varepsilon_v - U_v)/D)$ which understandably remain unchanged as there is no hybridization between the Fermi sea and the valence band. The next two correspond to $((\varepsilon_c + U_c)/D)$ and $(\varepsilon_c /D)$ with separation(or the renormalized coulomb gap $\Delta$) equal to 0.3378. The root $z_5$ corresponds to the the occurrence of DOS peak close to Fermi level. We find that a given hybridization parameter $V \ll D$ between localized and extended states leads to marginal enhancement in the coulomb gap.

We now make use of Schrieffer-Wolff transformation [44] to obtain a Kondo Hamiltonian [45,46], consisting of a unpaired spin S interacting locally with a non-interacting conduction electron sea ($H_{Kondo} = (JS/2)\sum_{\sigma,\sigma'}\psi^\dagger_\sigma \tau_{\sigma,\sigma'} \psi_{\sigma'} + H_{cond}$ where J denotes the Kondo coupling, $\psi^\dagger_\sigma$ creates a conduction electron at the impurity site, and $\tau_{\sigma,\sigma'}$ is the vector of Pauli spin matrices), to discuss in brief some low-energy properties of the Anderson model. The term, $H_{cond}$, describes the conduction electron bath. We find that the coupling $J = [(V^2 U_c)/(|\varepsilon_c||\varepsilon_c + U_c|)]$. We show that $\log_e(D/T_k)$ ( where $T_k$ Kondo temperature which corresponds to the frequency at which the spin spectral function takes its maximum), given by Haldane's expression[47]

$$\log_e(D/T_k) = [\{\{2|(\varepsilon_c /D)| |((\varepsilon_c + U_c)/D)|\}/\{(V/D)^2 (U_c /D)\}] \qquad (11)$$

as a function of $a = |(\varepsilon_c /D)|$, reaches its maximum value at $|(\varepsilon_c /D)| = 0.1667$ and decreases with increasing $|(\varepsilon_c /D)|$ at a given V(see Table 1 where we have taken $V = D/18$). Beyond the value $|(\varepsilon_c /D)| = 0.1667$ we obtain complex roots of $(D_0 / D^5) = 0$ which is inconsistent with HFA framework as these roots correspond to excitations with finite lifetime. In HFA, since the Hamiltonian is completely diagonalized, the excitations are long-lived ones. We have also demonstrated in Table 1 that the renormalized Mott gap $\Delta$ increases with decrease in $\log_e(D/T_k)$. In fact, a second order curve-fitting for the range $3.0653 \leq \log_e(D/T_k) \leq 54$ yields $\Delta \approx 0.3984 - 0.0003 \log_e(D/T_k) - 0.00001 (\log_e(D/T_k))^2$.

We shall now show that the gap could be tuned due to the involvement of the dipole moment **D** of QD transitions. We take into account the coupling $\mu$ involving terms for this purpose. We further assume that the averages $\{\langle c^\dagger_\uparrow v_\downarrow\rangle, \langle c^\dagger_\downarrow v_\uparrow\rangle,...\}$ are quite small compared to unity in order to see how a token presence of the coupling $\mu$ will affect the v-bands $\{(\varepsilon_v /D),((\varepsilon_v - U_v)/D)\}$ and the c-bands $\{(\varepsilon_c /D),((\varepsilon_c + U_c)/D)\}$. This effectively means that we have not yet taken into account in full measure the excitation and the annihilation of excitons in the QD by photon absorption and photon emission respectively. Also, we have considered here the limiting case of vanishing exciton binding energy( $U_{coul} = 0$).We find that the counterpart of Eq. (9) now appears as

$$(z+1)[z^4 + (2a) z^3 + \{a^2 - (V/D)^2 + (a/3) - (1/9) - (\mu/D)^2\} z^2$$
$$+\{a((a/3) - (1/9) - (\mu/D)^2)) - a(V/D)^2 - (2/3)(V/D)^2\}z - (V/D)^2((1/9) + (a/3))] \approx 0. \qquad (12)$$

Obviously enough, though $((\varepsilon_v - U_v)/D)$ remains unchanged, $(\varepsilon_v/D)$ does get affected along with $\{(\varepsilon_c/D), ((\varepsilon_c+U_c)/D)\}$. We consider the quartic part within the square bracket in (12). The roots and the results obtained are summarized in Table 2(A). We have shown here, as before, that the renormalized Mott gap $\Delta$ increases with decrease in $\log_e(D/T_k)$. In fact, a second order curve-fitting for the range $3.0653 \leq \log_e(D/T_k) \leq 54$ yields $\Delta \approx 0.4040 - 0.0005 \log_e(D/T_k) - 0.00001 (\log_e(D/T_k))^2$. We find that the involvement of the dipole moment of the QD transition enhances $\Delta$ in comparison with the case when the involvement is absent.

The single-dot spectral function (SF) in the c-band spin-$\sigma$ channel ( i.e. the one corresponding to $G_c^{\sigma,\sigma}(\tau) = -\langle T\{c_\sigma(\tau) c^\dagger_\sigma(0)\}\rangle$ ) is given by $A_c^{\sigma,\sigma}(\omega) = (-\pi^{-1})\text{Im } G_c^{(R)\sigma,\sigma}(\omega)$, where $G_c^{(R)\sigma,\sigma}(\omega)$ is a retarded Green's function given by

$$G_c^{(R)\sigma,\sigma}(\omega) = {}_{-\infty}\!\int^{\infty} (d\omega'/2\pi)\{\zeta_c^{(R)\sigma,\sigma}(\omega')/(\omega - \omega' + i\,0^+)\} \tag{13}$$

and $\zeta_c^{(R)\sigma,\sigma}(\omega) = -(i/2\pi)\{G_c^{(R)\sigma,\sigma}(z)|_{z=\omega-i0+} - G_c^{(R)\sigma,\sigma}(z)|_{z=\omega+i0+}\}$. The spectral functions provide information about the nature of the allowed electronic states, regardless whether they are occupied or not. In Eq.(12) we set $(\mu/D)=(1/18)$ and allow V to vary. With the aid of roots of (12)(see Table 2(B)), the corresponding coherence factors and Eq.(13) we express the c-band spin-$\sigma$ channel spectral function which is found to be a bunch of delta functions. Here we notice that , as a function of the tuning parameter $v = (V/D)$, the system undergoes a crossover from the state where the conduction band and the c-level are fully coupled to a state where these are decoupled. To clarify, for $v > v_c \to 0^+$ ($v_c \approx 0.01$) the system is in a phase where the c- electrons present a Kondo peak with infinite lifetime and reasonable peak height close to the Fermi level and take active part in the conduction. The c-electron spectrum, however, has a gap $\Delta$ together with the presence of the Kondo peak(see Fig.2) . For $v < v_c$ we have a Mott-like localization, where the c-electron spectrum again has a gap $\Delta$ without the Kondo peak(see Fig.2). Within this gap the conduction electrons fully recover the free band DOS and the effective hybridization is practically zero. This shows that the conduction band at low energy is completely decoupled from the c-level, but the effective hybridization remains active at a finite intermediate energy scale. It is also noticed that starting from $v_c$ as the tuning parameter v is increased there is an overall decrease in the Mott gap $\Delta$. Whether this crossover is actually a phase transition or not could be answered through an investigation based on a superior framework such as Hubbard approximation, Dynamic mean field theory ( DMFT), etc.. This is a future task. It must be noted that v cannot be as high as unity here, for then we shall have complex roots of Eq.(12) which are not possible, as had already been explained above, in HF approximation scheme.

We next consider a local retarded Green's function $G^R_{c\sigma}(\omega) = (-i) {}_{-\infty}\!\int^{\infty} dt \exp(i\omega t)\langle\{c_\sigma(t), c^\dagger_\sigma(0)\}\rangle\theta(t)$. The local density of states (LDOS) in the c-band spin-$\sigma$ channel ( $\rho_c^{\sigma,\sigma}(\omega)$) is given by $\rho_c^{\sigma,\sigma}(\omega) = (-\pi^{-1})\text{Im } G^R_{c\sigma}(\omega)$. We find, in units such that $\hbar = 1$, that LDOS $\rho_c^{\sigma,\sigma}(\omega) = \tanh(\beta\omega/2) A_c^{\sigma,\sigma}(\omega)$. Upon using the result $(x \pm i\,0^+)^{-1} = [P(x^{-1}) \pm (1/i)$ π δ (x)], where P represents a Cauchy's principal value, in view of (12) we find that the spin-up channel LDOS $\rho_c^{\sigma,\sigma}(\omega)$ is given by a bunch of delta functions ( a Fermi-liquid-like feature) in the form $\rho_c^{\sigma,\sigma}(\omega/D) \approx \tanh(\beta\omega/2) [\sum_{j=1,2,3,4} A_c^{(j)\uparrow\uparrow} \delta((\omega/D) - e_j) + A_c^{(5)\uparrow\uparrow} \delta((\omega/D) + 1)]$ where ($e_1, e_2, e_3, e_4$) are the four roots of the quartic part within the square bracket in (12) and given in Table 2(A) for different values of $\log_e(D/T_k)$. Since the coefficients ($A_c^{(j)\uparrow\uparrow}, A_c^{(5)\uparrow\uparrow}$) are found to be non-zero, all the five states $e = e_j$ and $e = -1$ are sufficiently long-lived ones. We now replace the delta functions above by the well-known result $\delta(x) = \lim_{\Gamma \to 0} (1/\pi)(\Gamma/(x^2 + \Gamma^2))$ to obtain a graphical representation of $\rho_c^{\sigma,\sigma}(\omega)$. In Fig.3(A) we have depicted, for T = 300 K and D = 10 meV, $\rho_c^{\sigma,\sigma}(\omega/D)$–vs –$(\omega/D)$ sketch for B= $\log_e(D/T_k)$ = 3.0653. The two numerical values B= $\log_e(D/T_k)$ = 54.0000 and B= $\log_e(D/T_k)$ = 3.0653 represent the two extremes of the local moment regime $-U_c + \Gamma \leq \varepsilon_c \leq \Gamma$. One finds that as $T_k$ increases, the dot behaving as a Kondo impurity and showing a peak ( at $e_2$) in the LDOS close to the Fermi level thereby leading to characteristic Kondo-Abrikosov-Suhl (KAS) resonances with enhanced conductance around zero bias, shifts towards slightly higher value of $|\omega/D|$. Furthermore, at a given temperature, as v = (V/D) tends towards $0^+$ the c-fermion DOS (or LDOS) show both a suppression of the peak close to Fermi level and a shift to less negative energies. The increase in temperature leads to the peak reduction in the KAS resonance(the peak-broadening does not occur as this is not possible in HFA framework). The effect is reminiscent of the de-coherence caused by the measurement leading to the pure decrease of the amplitude of the resonance without any smearing. We shall show in the next section that a decrease in temperature gives rise to absorption peak reduction.

### III. SYMMETRISED AUTOCORRELATION FUNCTION

The Fourier coefficient $G_c(\uparrow, z)$ and $G_v(\uparrow, z)$ lead to spectral function (and LDOS). With these one obtains the effect of hybridization parameter and the dipole moment of QD transition on the single particle spectrum ( and conductance around zero bias) . As already stated, our aim is to investigate the optical absorption/emission (A/E) coefficients. In a bid to do this, a symmetrized quantum autocorrelation function, which gives the frequency dependent transition rate corresponding to the absorption and emission phenomena, is introduced below. Generally speaking, for a quantum system subjected to a time-dependent external driving field μ(t) ( = μ exp(iωt)), such that the full Hamiltonian takes the form H = ($H_1$ + μ(t) Â) where Â is an operator through which coupling occurs, the optical transitions between eigen-states of $H_1$ could be induced by the field. Consider such a transition between an initial state | i › and a final state | f › of $H_1$ with eigen-energies $E_i$ and $E_f$ respectively, where $E_f = E_i + \hbar\omega$. The transition rate can actually be determined using the Fermi Golden Rule which states that the probability of a transition occurring per unit time, $R_{i \to f}$, is given by $R_{i \to f}(\omega) = (2\pi/\hbar) |\langle f| \mu \hat{A} |i \rangle|^2 \partial(E_f - (E_i + \hbar\omega))$; the delta function expresses the fact that energy is conserved. The net transition rate is obtained by summing over both i and f and weighing the sum by the probability $P(\omega) = \sum_{i,f} R_{i \to f}(\omega) \rho_i$ such that the initial state of the system is | i ›. Here $\rho_i$ is an eigenvalue of the density matrix [ exp(−β$E_i$)/ Tr ( exp(−βH)], where β = (kT)$^{-1}$. This leads to the transition rates corresponding to absorption and emission, respectively, as $P(\omega) = (2\pi/\hbar) |\mu|^2 C_>(\omega)$ and $P(-\omega) = (2\pi/\hbar) |\mu|^2 C_<(\omega)$ where $C_>(\omega) = \sum_{i,f} \rho_i |\langle f|\mu \hat{A}|i\rangle|^2 \partial(E_f - (E_i + \hbar\omega))$ and $C_<(\omega) = \sum_{i,f} \rho_i |\langle f|\mu \hat{A}|i\rangle|^2 \partial(E_f - (E_i - \hbar\omega))$. It is easy to see that $C_<(\omega) = \exp(-\beta \hbar\omega) C_>(\omega)$ which gives $P(-\omega) = \exp(-\beta \hbar\omega) P(\omega)$. This is the equation of detailed balance. We see from it that the probability of emission is less than that for absorption. The reason for this is that it is less likely to find the system in an excited state $|f\rangle$ initially, when it is in contact with a heat bath and hence thermally equilibriated. Note that, though $R_{i \to f}(\omega) = R_{i \to f}(-\omega)$ (i.e. microscopic laws of motion are reversible), we have $P(\omega) > P(-\omega)$. The conclusion is that reversibility is lost when the system is in contact with a heat bath.

Upon using the representations $\partial(E) = (2\pi)^{-1} \int_{-\infty}^{\infty} dt \exp(-i\omega t)$ and $\hat{A}(t) = \exp(i H_1 t) \hat{A} \exp(-i H_1 t)$ we find that $C_>(\omega)$ and $C_<(\omega)$ may be expressed in terms of the statistical averages $\langle \hat{A}(t)\hat{A}(0)\rangle$ and $\langle \hat{A}(0)\hat{A}(t)\rangle$ involving the density matrix alluded to above. We also find

$$(C_>(\omega) - C_<(\omega)) = (C_>(\omega) + C_<(\omega))\{(1 - \exp(-\beta \hbar\omega))/(1 + \exp(-\beta \hbar\omega))\}$$

which leads to the following expression for the energy difference $Q(\omega) = \hbar\omega[P(\omega) - P(-\omega)]$:

$$Q(\omega) = (\omega/\hbar) |\mu|^2 \tan(\beta \hbar\omega/2) \int_{-\infty}^{\infty} dt \exp(i\omega t) \langle\{\hat{A}(0), \hat{A}(t)\}\rangle. \tag{14}$$

The quantity $\langle\{\hat{A}(0), \hat{A}(t)\}\rangle$ is the symmetrized quantum autocorrelation function (SQAF) mentioned above. The frequency dependent transition rates corresponding to absorption and emission, respectively, on the other hand are given by

$$P_{absorption}(\omega) = \hbar^{-2} |\mu|^2 \int_{-\infty}^{\infty} dt \exp(i\omega t) \langle \hat{A}(t)\hat{A}(0)\rangle,$$

$$P_{emission}(\omega) = \hbar^{-2} |\mu|^2 \int_{-\infty}^{\infty} dt \exp(i\omega t) \langle \hat{A}(0)\hat{A}(t)\rangle. \tag{15}$$

The frequency spectrum corresponding to the SQAF is defined as $G(\omega) = (1/2\pi) \int_{-\infty}^{\infty} dt \exp(i\omega t) \langle(1/2)\{\hat{A}(0), \hat{A}(t)\}\rangle$. Our task now is to express $\langle \hat{A}(t)\hat{A}(0)\rangle$ and $\langle \hat{A}(0)\hat{A}(t)\rangle$ in terms of a temperature Green's function to facilitate explicit calculation of $P_{absorption}(\omega)$ and $P_{emission}(\omega)$.

We consider the temperature function $D(\tau, \tau') = -\langle T\{\hat{A}(\tau) \hat{A}(\tau')\}\rangle$ for the purpose stated above, where as before T is the time-ordering operator which arranges the operators from right to left in the ascending order of imaginary time $\tau$. We write, for $\tau > 0$, $D(\tau, 0) = -\langle \hat{A}(\tau) \hat{A}(0)\rangle$. Here the trace can be evaluated in any basis. A particularly convenient choice is the exact eigen-state of H: $H|m\rangle = H_m |m\rangle$. The Lehmann representation of the Fourier coefficient $D(i\omega_n)$ of $D(\tau, 0)$ is then given by $D(i\omega_n) = \exp(\beta Ж) \sum_{mn} \exp(-\beta H_m) \langle m|\hat{A}|n\rangle\langle n|\hat{A}|m\rangle\{(1 - \exp(-\beta(H_n - H_m)))/(i\omega_n - (H_n - H_m))\}$ where Ж is the thermodynamic potential of the system. With the aid of this representation it is straightforward to show that

$$\langle \hat{A}(0)\hat{A}(t)\rangle = (-\pi^{-1}) \exp(-\beta \hbar\omega) \int_{-\infty}^{\infty} d\omega' \exp(-i\omega' t)$$

$$(1- \exp(-\beta \hbar \omega'))^{-1} \operatorname{Im} (D(i\omega_n) \big|_{i\omega n = \omega' + i0+}) \qquad (16)$$

where $i\omega_n$ are Matsubara frequencies. The quantity $\langle \hat{A}(t)\hat{A}(0)\rangle = \exp(\beta \hbar \omega) \langle \hat{A}(0)\hat{A}(t)\rangle$. For the problem at hand, as it is clear from Eq.(4), the operator $\hat{A} = \Sigma_\sigma (c^\dagger_\sigma v_\sigma + v^\dagger_\sigma c_\sigma)$. It follows that formally one can write $\operatorname{Im}(D(i\omega_n)\big|_{i\omega n = \omega' + i0+})$ as

$$\operatorname{Im}(D(i\omega_n)\big|_{i\omega n = \omega' + i0+}) = -\pi \exp(\beta \text{Ж}) \sum_{mn} \exp(-\beta H_m) \langle m|\hat{A}|n\rangle \langle n|\hat{A}|m\rangle$$

$$\{1- \exp(-\beta (H_n - H_m))\} \partial(\omega' - (H_n - H_m)). \qquad (17)$$

In units such that $\hbar = 1$, in view of (16) and (17), one can write

$$P_{absorption}(\omega) = |\mu|^2 \int_{-\infty}^{\infty} d\omega' \exp(\beta \text{Ж}) \int_{-\infty}^{\infty} dt \exp(i(\omega - \omega')t)$$

$$\sum_{mn} \exp(-\beta H_m) \langle m|\hat{A}|n\rangle \langle n|\hat{A}|m\rangle \partial(\omega' - (H_n - H_m)). \qquad (18)$$

As regards the frequency dependent transition rate corresponding to emission, it is given by a similar expression; an additional multiplicative factor $\exp(-\beta \omega)$ will appear. One might expect that the absorption and emission spectrum are related by the particle-hole symmetry (PHS). To explain, consider the operators $J^+ \equiv c^\dagger_\uparrow c^\dagger_\downarrow$, $J^- = (J^+)^\dagger$ and $J_z = [J^+, J^-]$ which form a SU(2) algebra, where $J_z = c^\dagger_\uparrow c_\uparrow + c^\dagger_\downarrow c_\downarrow$ is proportional to the charge operator. The particle-hole symmetry exists if $J^+$ commutes with the Hamiltonian. However, the Hamitonian under consideration has no PHS, as with $H_1 \approx H_0 + H_I$ given by (5) one obtains

$$[H_1, J^+] = (\varepsilon_c - U_{coul} + U_c) c^\dagger_\uparrow c^\dagger_\downarrow + \Sigma_k V_k d^\dagger_{k\uparrow} c^\dagger_\downarrow + \beta v^\dagger_\uparrow c^\dagger_\downarrow$$

$$+ \Delta_c c^\dagger_\downarrow c^\dagger_\downarrow + (\Delta^\dagger_c + U_c) c^\dagger_\uparrow c^\dagger_\uparrow + \Delta_1 v^\dagger_\downarrow c^\dagger_\downarrow, \qquad (19)$$

and therefore a relation between absorption and emission spectrum by PHS is denied. From Eq. (18) one can see that the absorption coefficient is only approximately proportional to $|\mu|^2 \int d\omega' \Sigma_{\sigma,\sigma'} \Omega_{\sigma,\sigma'}(\omega', \omega - \omega')$ where $\Omega_{\sigma,\sigma'}(\omega', \omega - \omega') = [(A_1^\sigma(\omega') + A_2^\sigma(\omega'))(A_1^{\sigma'}(\omega-\omega') + A_2^{\sigma'}(\omega-\omega'))]$ and $A_1^\sigma(\omega)$ and $A_2^\sigma(\omega)$, respectively, are the spectral functions corresponding to the Green's functions $G_{vc}(\sigma, \tau) = -\langle T\{v_\sigma(\tau) c^\dagger_\sigma(0)\}\rangle$ and $G_{cv}(\sigma, \tau) = -\langle T\{c_\sigma(\tau) v^\dagger_\sigma(0)\}\rangle$. It has been argued in the literature [48,49,50], that the convolution of electron and hole spectral functions gives a reasonable approximation to the absorption spectrum. One notices here the approximate nature of such an approach. As already stated, we shall, however, be following EOM method as it is simple and reliable given the Hartree-Fock model Hamiltonian (5). We would like to point out that the absorption spectrum of dot consists of delta peaks only. This was clear from the beginning since a finite perturbation cannot change the character of the spectrum of the unperturbed system. Therefore, our approach too is expected to yield the same with slightly varying details.

We wish to evaluate the temperature function $D(\tau, 0) = -\Sigma_{\sigma, \sigma'} \langle T \{ (c^{\dagger}_{\sigma}(\tau)v_{\sigma}(\tau) + v^{\dagger}_{\sigma}(\tau)c_{\sigma}(\tau))(c^{\dagger}_{\sigma'}(0)v_{\sigma'}(0) + v^{\dagger}_{\sigma'}(0)c_{\sigma'}(0)) \} \rangle$. This actually requires calculating four temperature functions

$$G_1(\tau, 0) = -\Sigma_{\sigma, \sigma'} \langle T \{ c^{\dagger}_{\sigma}(\tau)v_{\sigma}(\tau) c^{\dagger}_{\sigma'}(0)v_{\sigma'}(0) \} \rangle,$$

$$G_2(\tau, 0) = -\Sigma_{\sigma, \sigma'} \langle T \{ c^{\dagger}_{\sigma}(\tau)v_{\sigma}(\tau) v^{\dagger}_{\sigma'}(0)c_{\sigma'}(0) \} \rangle,$$

$$G_3(\tau, 0) = -\Sigma_{\sigma, \sigma'} \langle T \{ v^{\dagger}_{\sigma}(\tau)c_{\sigma}(\tau) c^{\dagger}_{\sigma'}(0)v_{\sigma'}(0) \} \rangle,$$

$$G_4(\tau, 0) = -\Sigma_{\sigma, \sigma'} \langle T \{ v^{\dagger}_{\sigma}(\tau)c_{\sigma}(\tau) v^{\dagger}_{\sigma'}(0)c_{\sigma'}(0) \} \rangle. \tag{20}$$

Take, for example, the first one $G_1(\tau, 0) = -\Sigma_{\sigma, \sigma'} \langle T \{ c^{\dagger}_{\sigma}(\tau)v_{\sigma}(\tau) c^{\dagger}_{\sigma'}(0)v_{\sigma'}(0) \} \rangle$. This comprises of four temperature functions as shown below:

$$G_1(\tau, 0) = G_1^{\uparrow\downarrow}(\tau, 0) + G_1^{\downarrow\uparrow}(\tau, 0) + G_1^{\uparrow\uparrow}(\tau, 0) + G_1^{\downarrow\downarrow}(\tau, 0),$$

$G_1^{\uparrow\downarrow}(\tau, 0) = -\langle T \{ c^{\dagger}_{\uparrow}(\tau)v_{\uparrow}(\tau) c^{\dagger}_{\downarrow}(0)v_{\downarrow}(0) \} \rangle$, $G_1^{\downarrow\uparrow}(\tau, 0) = -\langle T \{ c^{\dagger}_{\downarrow}(\tau)v_{\downarrow}(\tau) c^{\dagger}_{\uparrow}(0)v_{\uparrow}(0) \} \rangle$,

$G_1^{\uparrow\uparrow}(\tau, 0) = -\langle T \{ c^{\dagger}_{\uparrow}(\tau)v_{\uparrow}(\tau) c^{\dagger}_{\uparrow}(0)v_{\uparrow}(0) \} \rangle$, $G_1^{\downarrow\downarrow}(\tau, 0) = -\langle T \{ c^{\dagger}_{\downarrow}(\tau)v_{\downarrow}(\tau) c^{\dagger}_{\downarrow}(0)v_{\downarrow}(0) \} \rangle$.

(21)

We shall have to find equations of motion of the four functions $\{ G_1^{\uparrow\uparrow}(\tau, 0), G_1^{\uparrow\downarrow}(\tau,0), G_1^{\downarrow\uparrow}(\tau, 0), G_1^{\downarrow\downarrow}(\tau, 0) \}$. A similar exercise has to be done for the functions $G_2(\tau, 0)$, $G_3(\tau, 0)$ and $G_4(\tau, 0)$. In fact, as we can see, we shall have to evaluate sixteen functions to get the final expression for $D(\tau,0)$ which will yield an expression for the autocorrelation functions $\langle \hat{A}(0)\hat{A}(t) \rangle$ and $\langle \hat{A}(t)\hat{A}(0) \rangle$. These functions will ultimately lead to the frequency dependent transition rates corresponding to absorption and emission given by Eq.(15).

We find the following equation for $G_1^{\uparrow\downarrow}(\tau, 0)$ for a single vector k ($|k| \approx |k_F|$, with $k_F$ the Fermi momentum): $(\partial/\partial\tau) G_1^{\uparrow\downarrow}(\tau, 0) \approx [(\varepsilon_c + U_c - \varepsilon_v) G_1^{\uparrow\downarrow}(\tau, 0) + V G_1^{(I)\uparrow\downarrow}(\tau, 0) + \beta G_1^{(II)\uparrow\downarrow}(\tau, 0)]$, where

$$G_1^{(I)\uparrow\downarrow}(\tau, 0) = -\langle T \{ d^{\dagger}_{k\uparrow}(\tau)v_{\uparrow}(\tau) c^{\dagger}_{\downarrow}(0)v_{\downarrow}(0) \} \rangle,$$

$$G_1^{(II)\uparrow\downarrow}(\tau, 0) = -\langle T \{ v^{\dagger}_{\uparrow}(\tau)v_{\uparrow}(\tau) c^{\dagger}_{\downarrow}(0)v_{\downarrow}(0) \} \rangle. \tag{22}$$

Similarly, the equation of motion of $G_1^{(I)\uparrow\downarrow}(\tau, 0)$ is $(\partial/\partial\tau) G_1^{(I)\uparrow\downarrow}(\tau, 0) \approx [(\varepsilon_k - h - \varepsilon_v) G_1^{(I)\uparrow\downarrow}(\tau, 0) + V G_1^{\uparrow\downarrow}(\tau, 0) - \beta G_1^{(III)\uparrow\downarrow}(\tau, 0)]$ where $G_1^{(III)\uparrow\downarrow}(\tau, 0) = -\langle T \{ d^{\dagger}_{k\uparrow}(\tau)c_{\uparrow}(\tau) c^{\dagger}_{\downarrow}(0)v_{\downarrow}(0) \} \rangle$. The equations of motion of the remaining functions $G_1^{(II)\uparrow\downarrow}(\tau, 0)$, $G_1^{(III)\uparrow\downarrow}(\tau, 0)$ and $G_1^{(IV)\downarrow\downarrow}(\tau, 0)$ are

$(\partial/\partial\tau) G_1^{(II)\uparrow\downarrow}(\tau, 0) = \beta G_1^{\uparrow\downarrow}(\tau, 0) - \beta G_1^{(III)\uparrow\downarrow}(\tau, 0) + \Delta_v G_1^{(IV)\downarrow\downarrow}(\tau, 0)$,

$(\partial/\partial\tau) G_1^{(III)\uparrow\downarrow}(\tau, 0) = (\varepsilon_k - h - \varepsilon_c - U_c) G_1^{(III)\uparrow\downarrow}(\tau, 0) - \beta G_1^{(I)\uparrow\downarrow}(\tau, 0)$,

$(\partial/\partial\tau) G_1^{(IV)\downarrow\downarrow}(\tau, 0) = (\varepsilon_v - U_v) G_1^{(IV)\downarrow\downarrow}(\tau, 0) + \Delta_2 G_1^{\uparrow\downarrow}(\tau, 0)$

$$+ (\Delta^\dagger_v + U_v) G_1^{(II)\uparrow\downarrow}(\tau, 0) + \langle c^\dagger_\downarrow v_\uparrow \rangle \partial(\tau). \quad (23)$$

The Fourier coefficients of the thermal averages above are the Matsubara propagators $\{G_1^{\uparrow\downarrow}(z'), G_1^{(I)\uparrow\downarrow}(z'), G_1^{(II)\uparrow\downarrow}(z'), G_1^{(III)\uparrow\downarrow}(z'), G_1^{(IV)\downarrow\downarrow}(z')\}$ where $z'$ is integer multiple of $(\pi i / \beta D)$. Assuming $(\mu/D, U_c/D,..) \ll 1$, in the Kondo regime, we find that the poles of the Fourier coefficient $G_j(z')$ correspond to

$$z'_1 = 0, \; z'_2 = (\varepsilon_c + U_c), \; z'_3 = [-((\varepsilon_c + U_c)/2 - \varepsilon_v) + \sqrt{\{((\varepsilon_c + U_c)/2)^2 + V^2\}}],$$

$$z'_4 = [-((\varepsilon_c + U_c)/2 - \varepsilon_v) - \sqrt{\{((\varepsilon_c + U_c)/2)^2 + V^2\}}], \; z'_5 = (\varepsilon_v - U_v). \quad (24)$$

The correlation between anti-parallel spin(densities) is not taken into consideration above. We now retain the coulomb interaction induced term $\Delta_v G_1^{(IV)\downarrow\downarrow}(\tau, 0)$ as states with two holes are very highly excited states and it will be interesting to observe whether the actual value of $U_v$ is likely to have influence on final results. With $(\mu/D)$ involving terms included, for the coupling strengths $(\mu/D) = 0.1667 = (V/D)$ and the other parameter values same as in Section 2, it is straightforward to see that

$$\dot{D} G_1^{\uparrow\downarrow}(z) \approx (\langle c^\dagger_\downarrow v_\uparrow \rangle/27)\{[a_1/(z - 0.0062)] + [a_2/(z - 0.1731)] + [a_3/(z + 0.3237)]$$

$$+ [a_4/(z + 0.5222] + [a_5/(z + 1.0000)]\},$$

$$DG_1^{\uparrow\uparrow}(z) \approx (\langle c^\dagger_\uparrow v_\uparrow \rangle/18)\{[b_1/(z - 0.0062)] + [b_2/(z - 0.1731)] + [b_3/(z + 0.3237)]$$

$$+ [b_4/(z + 0.5222)] + + [b_5/(z + 1.0000)]\}, \quad (25)$$

and so on. The numerical factors $(a_1, a_2, ....)$ have been calculated solving Eqs.(22) and (23) explicitly. We find, in view of (15), the dimensionless energy values at which peaks in the absorption/emission coefficients are expected to occur are $\{0.0062, 0.1731, -0.3237, -0.5222, -1.000\}$. These results lead to the expressions

$$P_{absorption}(\omega/D) = |\mu|^2 \int_{-\infty}^{\infty} dt \exp(i\omega t) \langle \hat{A}(t)\hat{A}(0)\rangle$$

$$= |\mu|^2 [C_1 (1 - \exp(-0.0062\beta D))^{-1} \partial((\omega/D) - 0.0062) + C_2 (1 - \exp(-0.1731\beta D))^{-1}$$

$$\partial((\omega/D) - 0.1731) + C_3 (1 - \exp(0.3237\beta D))^{-1} \partial((\omega/D) + 0.3237)$$

$$+ C_4 (1 - \exp(0.5222\beta D))^{-1} \partial((\omega/D) + 0.5222)$$

$$+ C_5 (1 - \exp(\beta D))^{-1} \partial((\omega/D) + 1.0000)] \quad (26)$$

where the coefficients ( $C_1$, $C_2$,……) have been determined, as in Section 2, setting up and solving EOMs of the Green's functions corresponding to the averages $\langle c^\dagger_\downarrow v_\uparrow \rangle$, $\langle c^\dagger_\uparrow v_\uparrow \rangle$, and so on.  It must be mentioned that we have considered here the  limiting case of vanishing exciton binding energy( $U_{coul} = 0$).

The quantity $P_{emission}(\omega) =  | \mu |^2 \int_{-\infty}^{\infty} dt \exp(i\omega t) \langle \hat{A}(0)\hat{A}(t)\rangle$, on the other hand, is given by

$$P_{emission}(\omega/D) = | \mu |^2 [ C_1 \exp(-0.0062\beta D ) (1- \exp(-0.0062\beta D ))^{-1} \partial ((\omega/D) - 0.0062)$$

$$+ C_2 \exp(-0.1731\beta D )(1- \exp(-0.1731\beta D ))^{-1} \partial ((\omega/D) - 0.1731)$$

$$+ C_3 \exp(0.3237\beta D ) (1- \exp(0.3237\beta D ))^{-1} \partial ((\omega/D) + 0.3237)$$

$$+ C_4 \exp(0.5222\beta D )(1- \exp(0.5222\beta D ))^{-1} \partial ((\omega/D) +0.5222)$$

$$+ C_5 \exp(\beta D ) (1- \exp(\beta D ))^{-1} \partial ((\omega/D) + 1.0000)]. \qquad (27)$$

The constants ($C_1$, $C_2$,....) have been determined self-consistently by EOM method. In the HFA framework adopted reasonable outcome of these details are assured. The linear absorption spectrum of the dot obtained from (26) and (27)is shown in Fig.4 for  300 K. The spectrum consists of a series of resonances. The series correspond to two discrete quantum confined electron and hole states we started with. The spectrum reveals a sharp, tall peak close to KAS peak (in Fig.3) and a few smaller, distinguishable peaks on either side. The former clearly indicates that the Kondo effect, which is hybridization (V) induced, does leave its imprint on the absorption and emission ( non-Kondo processes), which are driven by the coupling $\mu$ involving the dipole moment of QD transitions reflecting the physical structure of the dot including the confinement potential, in the Kondo regime. The coupling $\mu$ determines the height of the absorption/emission peaks. In the weak coupling regime (($\mu/D) \ll 1$),  we obtain reduced peaks whereas in the strong coupling regime (($\mu/D) \sim 1$) taller peaks. A decrease in temperature gives rise to peak reduction in both the cases. The actual value of $U_v$  has no major influence on final results. We see an important feature: There is a threshold energy $\omega_0$(the energy below which no photon is absorbed/ emitted) and the threshold  $\omega_0$ shows a shift as a function of the exciton binding energy $U_{coul}$ . This is depicted in Fig.5 for the absorption spectrum with  $U_{coul} = D/6$.

## IV. DISCUSSION

In this communication, we have not taken into account the scattering of either electrons or holes by phonons. Even though a scattering event ( describable by a Hamiltonian $H_{phonon-e/h} = \Sigma_{k,\sigma, \sigma'} [\{M^{(e)}_{k,\sigma, \sigma'} c^\dagger_\sigma c_{\sigma'} + M^{(h)}_{k,\sigma, \sigma'} v_\sigma v^\dagger_{\sigma'} \}( b_k + b^\dagger_k )]$ where $b^\dagger_k$ creates a phonon with wave vector **k**  and frequency $\omega_k$ and the scattering matrix elements $M^{(e)}_{k,\sigma, \sigma'}$ depends on the type of phonons involved) changes the state of an electron or hole, the final scattered state may remain an optically active electron-hole bound state (exciton). Since such states play an important role in optical spectra, their inclusion seems to be necessary for a comprehensive investigation.

A quantum dot array possesses electronic states that go far beyond the 'artificial molecule' model. These states are a coherent hybridization of localized dot states and extended continuum states: they have no analogue in atomic and molecular physics. In a realistic device, such as a quantum dot solar cell ( QDSC), we have a dot array. To investigate the optical absorption aspect, which influences the short-circuit current of this system, we shall have to extend the model considered here to include the direct hopping of itinerant electrons between nearest neighbor (NN) dots/impurities. One may introduce the term $[-t\sum_{(ij)} (c^{\dagger}_i c_j + H.c.)]$, as has been done in Ref.[51] to investigate spin-pairing and inducement of semi-metallic state, in the Hamiltonian above where t is the tunneling coupling parameter between the NN dots and cast it as a multi-impurity Anderson problem. We believe that a theoretical investigation of QDSC in this manner can only provide a reliable estimate of the required number of dots in the array, for the Kondo regime, mixed-valence regime,etc., to optimize the power conversion efficiency. This is a future task to be undertaken in the numerical renormalization group framework.

# REFERENCES

[1] K. Karrai, R. J. Warburton, C. Schulhauser , A. Högele, B. Urbaszek, E. J. McGhee, A.O. Govorov, J. M. Garcia, B. D. Gerardot and P. M. Petroff, Nature (London) 427, 135 (2004).

[2] R. J.Warburton, C. Sch¨aflein, D. Haft, F. Bickel, A. Lorke, K. Karrai, J. M. Garcia, W. Schoenfeld, and P. M. Petroff, Nature 405, 926 (2000).

[3] P. W. Anderson, Phys. Rev. 124, 41 (1961).

[4] A. Zrenner, E. Beham, S. Stufler, F. Findeis, M. Bichler, and G. Abstreiter, Nature (London) **418**, 612 (2002).

[5] P. Michler, A. Kiraz, C. Becher, W. V. Schoenfeld, P. M. Petroff, L. Zhang, E. Hu, and A. Imamoˇglu, Science **290**, 2282 (2000).

[6] E. Moreau, I. Robert, J. M. G´erard, I. Abram, L. Manin, and V.Thierry-Mieg, Appl. Phys. Lett. **79**, 2865 (2001).

[7] T. D. Happ, I. I. Tartakovskii, V. D. Kulakovskii, J. P. Reithmaier, M. Kamp, and A. Forchel, Phys. Rev. B **66**, 041303(R) (2002).

[8] J. P. Reithmaier, G. Sek, A. L¨offler, C. Hofmann, S. Kuhn, S. Reitzenstein, L. V. Keldysh, V. D. Kulakovskii, T. L. Reinecke, and A. Forchel, Nature (London) **432**, 197 (2004).

[9] T. Yoshie, A. Scherer, J. Hendrickson, G. Khitrova, H. M. Gibbs, G. Rupper, C. Ell, O. B. Shchekin, and D. G. Deppe, Nature (London) **432**, 200 (2004).

[10] D. Goldhaber-Gordon et al., Nature 391, 156 (1998).

[11] S. M. Cronenwett, T. H. Oosterkamp, and L. P. Kouwenhoven, Science 281, 540 (1998).

[12] J. Nyg°ard, D. H. Cobden, and P. E. Lindelof, Nature 408, 342 (2000).

[13] M. R. Buitelaar, T. Nussbaumer, and C. Sch¨onenberger, Phys. Rev. Lett. 89, 256801 (2002).

[14]. A. Kaminski, Yu. N. Nazarov, and L. I. Glazman, Phys.Rev. Lett. 83, 384 (1999).

[15] R. L´opez, R. Aguado, G. Platero, and C. Tejedor, Phys. Rev. B 64, 075319 (2001)).

[16] A. Thranhardt, C. Ell, G. Khitrova, and H.M.Gibbs, Phys.Rev.B 65,035327 (2002).

[17] A. Muller, Q.Q.Wang, P. Bianucci, C.K. Shih, and Q.K. Xue, cond-mat 0404398.

[18] T. H. Stievater *et al.*, Phys. Rev. Lett. **87**, 133603 (2001).

[19] H. Kamada, H. Gotoh, J. Temmyo, T. Takagahara, and H.Ando, Phys. Rev. Lett. 87, 246401 (2001).

[20] H. Htoon et al., Phys. Rev. Lett **88**, 087401 (2002).

[21] A. Zrenner et al., Nature **418**, 612 (2002).

[22] E. V. Baklanov, V. P. Chebotaev, Sov. Phys. JETP **34**, 490 (1972).

[23] B. R. Mollow, Phys. Rev. A. **5**, 2217 (1972).

[24 ] S. Haroche, F.Hartmann, Phys. Rev. A **6**, 1280 (1972).

[25] F. Y. Wu, S. Ezekiel, M. Ducloy, B. R. Mollow, Phys. Rev.Lett. **38**, 1077 (1977).

[26] M. T. Gruneisen, K. R. MacDonald, R. W. Boyd, J. Opt.Soc. Am. B **5**, 123 (1988).

[27] Xiaodong Xu *et al.*, Science **317**, 929 2007.

[28] A. Muller *et al.*, Phys. Rev. Lett. **99**, 187402 (2007).

[29] T. Unold *et al.*, Phys. Rev. Lett. **92**, 157401 (2004).

[30] G. Wrigge, I. Gerhardt, J. Hwang, G. Zumofen and V. Sandoghdar, Nature Phys. **4**, 60 (2008).

[31] W. Kohn and L.J. Sham, Phys.Rev.140, A 1133 (1965).

[32] M. Koskinen, M. Manninen, and S.M.Reimann, Phys.Rev.Lett.79,1389(1997).

[33] I.H. Lee, V.Rao, R.M.Martin, and J.P.Leburton, Phys.Rev.B 57,9035(1998).

[34] Y.Meir , N.S.Wingreen,and P.A. Lee,Phys.Rev. Lett. 66,3048(1991).

[35] Q.-f.Sun,H.Guo,and T.-h.Lin,Phys.Rev.Lett.87, 176601(2001).

[36] P.A. Maksym and T. chakraborty, Phys. Rev. Lett. 70, 485 (1993).

[37] J.J. Palacios, L.M. Moreno, G. Chiappe, E. Louis, and C. Tejedor, Phys. Rev. B 50,5760 ( 1994).

[38] M. Fujito, A. Natori, and H. Yasunaga, Phys.Rev. B 53,9952 (1996).

[39] H.-M. Muller and S.E.Konin, Phys.Rev. B 54, 14532 (1996).

[40] H. R. Krishnamurthy, J. W. Wilkins, and K. G. Wilson, Phys. Rev. B 21, 1003 (1980).

[41] H. R. Krishnaurthy, J. W. Wilkins, and K. G. Wilson, Phys. Rev. B 21, 1044 (1980).

[42]  J. Kondo, Solid State Phys. 23, 183 (1969).

[43]  C. Kittel, Quantum Theory of Solids (Wiley, New York, 1963).

[44] J. R. Schrieffer and P. A. Wolff, Phys. Rev. 149, 491 (1966).

[45] Mikio Eto and Yuli V. Nazarov, cond-mat/0002010.

[46] J. Martinek, Y. Utsumi, H. Imamura,J. Barna´s, S. Maekawa, J. K¨onig, and G. Sch¨on, cond-mat/0210006.

[47] R. ˇZitko and J. Bonˇca, cond-mat/0604336.

[48] K.Kral and Z.Khas, Phys.Rev.B 57 R2061(1998).

[49] T. Stauber, R. Zimmermann, and H. Castella, Phys. Rev. B 62, 7336 (2000).

[50] M. I. Vasilevskiy and E. V. Anda, and S. S. Makler, Phys. Rev. B 70, 035318 (2004); M.I. Vasilevskiy, R.P. Miranda, E.V. Anda, and S.S. Makler, Semicond. Sci. Technol. 19, S312 (2004).

[51] Partha Goswami and Avinashi Kapoor, Physica E (2008)(article in press), doi:10.1016/j.physe.2008.08.058.

**TABLE 1: Tabulation of $\log_e (D/T_k)$ –vs- $\Delta_{DM=0}$.**

| a | $\log_e (D/T_k)$ | $\Delta_{DM=0}$ |
|---|---|---|
| 0.1667 | 53.9892 | 0.3379 |
| 0.1800 | 53.6427 | 0.3426 |
| 0.2000 | 51.8270 | 0.3499 |
| 0.2200 | 48.4561 | 0.3550 |
| 0.2400 | 43.5300 | 0.3612 |
| 0.2500 | 40.4833 | 0.3642 |
| 0.2600 | 37.0488 | 0.3678 |
| 0.2700 | 33.2249 | 0.3710 |
| 0.2778 | 29.9724 | 0.3737 |
| 0.2900 | 24.4108 | 0.3786 |
| 0.3000 | 19.4206 | 0.3824 |
| 0.3100 | 14.0415 | 0.3869 |
| 0.3200 | 8.2737 | 0.3916 |
| 0.3285 | 3.0653 | 0.3956 |

**TABLE 2(A): Tabulation of $\log_e (D/T_k)$ –vs- $\Delta_{DM\neq 0}$ (= $(e_1 - e_4)$)**

| a | Four roots $(e_1, e_2, e_3, e_4)$ of the quartic part within the square bracket in Eq.(12). | B = $\log_e (D/T_k)$ | $\Delta_{DM\neq 0}$ = $(e_1 - e_4)$ |
|---|---|---|---|
| 0.1667 | 0.2016,−0.0526,−0.3393,−0.1431 | 54.0000 | 0.3447 |
| 0.1800 | 0.1901,−0.0514,−0.3395,−0.1592 | 53.6427 | 0.3493 |
| 0.1900 | 0.1816,−0.0511,−0.3396,−0.1709 | 52.9293 | 0.3525 |
| 0.1950 | 0.1774,−0.0511,−0.3396,−0.1766 | 52.4268 | 0.3530 |
| 0.1975 | 0.1753,−0.0511,-0.3397,-0.1795 | 52.1391 | 0.3548 |
| 0.1990 | 0.1740,-0.0511,-0.3397, -0.1812 | 51.9548 | 0.3552 |
| 0.2000 | | 51.8270 | 0.3555 |

| (V/D) | Four roots (e₁, e₂, e₃, e₄) | (V/D)² | $\Delta_{DM\neq 0}$ |
|---|---|---|---|
| | 0.1732,−0.0512,−0.3397,−0.1823 | | |
| 0.2050 | | 51.1301 | 0.3570 |
| | 0.1690,−0.0513,−0.3397,−0.1880 | | |
| 0.2100 | | 50.3360 | 0.3584 |
| | 0.1649 -0.0515, -0.3398,-0.1935 | | |
| 0.2200 | | 48.4561 | 0.3613 |
| | 0.1567,−0.0522,−0.3399,−0.2046 | | |
| 0.2400 | | 43.5300 | 0.3673 |
| | 0.1409,−0.0544,−0.3401,−0.2264 | | |
| 0.2500 | | 40.4833 | 0.3705 |
| | 0.1333,−0.0558,−0.3402,−0.2372 | | |
| 0.2600 | | 37.0488 | 0.3738 |
| | 0.1258,−0.0575,−0.3403,−0.2480 | | |
| 0.2700 | | 33.2249 | 0.3772 |
| | 0.1185,-0.0595, -0.3404, -0.2587 | | |
| 0.2778 | | 29.9724 | 0.3794 |
| | 0.1124,−0.0579,−0.3332,−0.2670 | | |
| 0.2900 | | 24.4108 | 0.3848 |
| | 0.1047,-0.0642,-0.3404,-0.2801 | | |
| 0.3000 | | 19.4206 | 0.3890 |
| | 0.0982,-0.0670,-0.3404,-0.2908 | | |
| 0.3100 | | 14.0415 | 0.3937 |
| | 0.0920,-0.0701,-0.3401,-0.3017 | | |
| 0.3200 | | 8.2737 | 0.3990 |
| | 0.0860,-0.0735,-0.3395,-0.3130 | | |
| 0.3285 | | 3.0653 | 0.4047 |
| | 0.0812,-0.0767,-0.3381,-0.3235 | | |

**TABLE 2(B): Tabulation of $v = (V/D)$ –vs- $\Delta_{DM\neq 0}$ (= $(e_1 - e_4)$ )**

| $(\beta/D) = 1/18$ | Four roots ($e_1$, $e_2$, $e_3$, $e_4$) of the quartic part within the square bracket in (17). | | $\Delta_{DM\neq 0}$ = ($e_1 - e_4$) |
|---|---|---|---|
| $\log_e (D/T_k) = 54.00$ | | | |
| (V/D), (V/D)² | | | |
| 0.001,0.000001 | | | 0.3400 |
| | 0.17284,−0.000017,−0.3390,−0.1672 | | |
| 0.0125,0.00015625 | | | 0.3402 |
| | 0.1745,−0.0026,−0.3395,−0.1657 | | |
| 0.0632,0.0040 | | | 0.3388 |
| | 0.2085,−0.0724,−0.3392,−0.1303 | | |

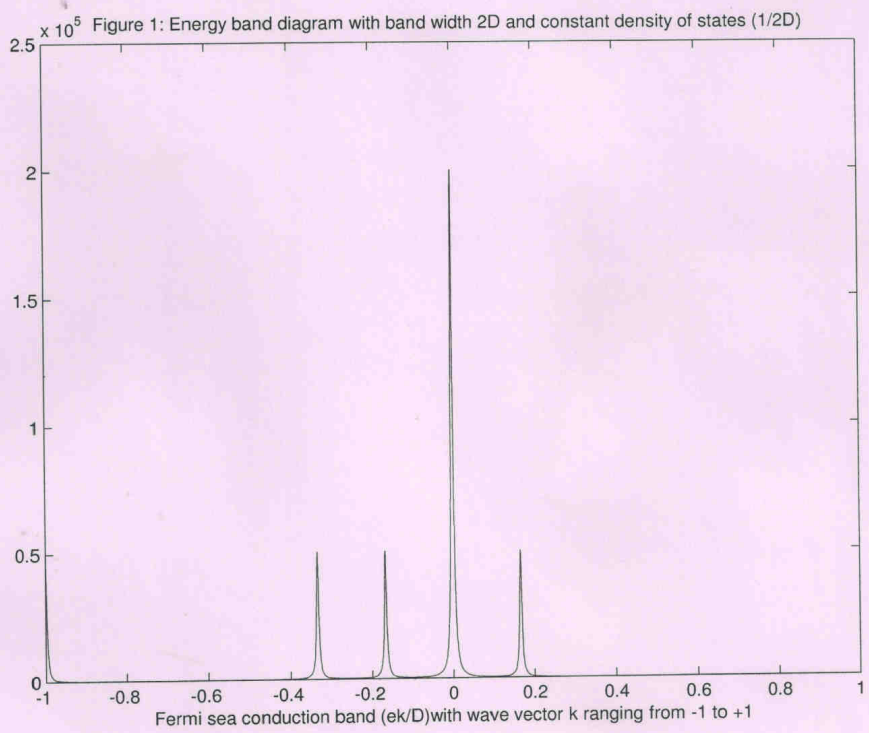

Figure 1: Energy band diagram with band width 2D and constant density of states (1/2D)

Fermi sea conduction band (ek/D) with wave vector k ranging from -1 to +1

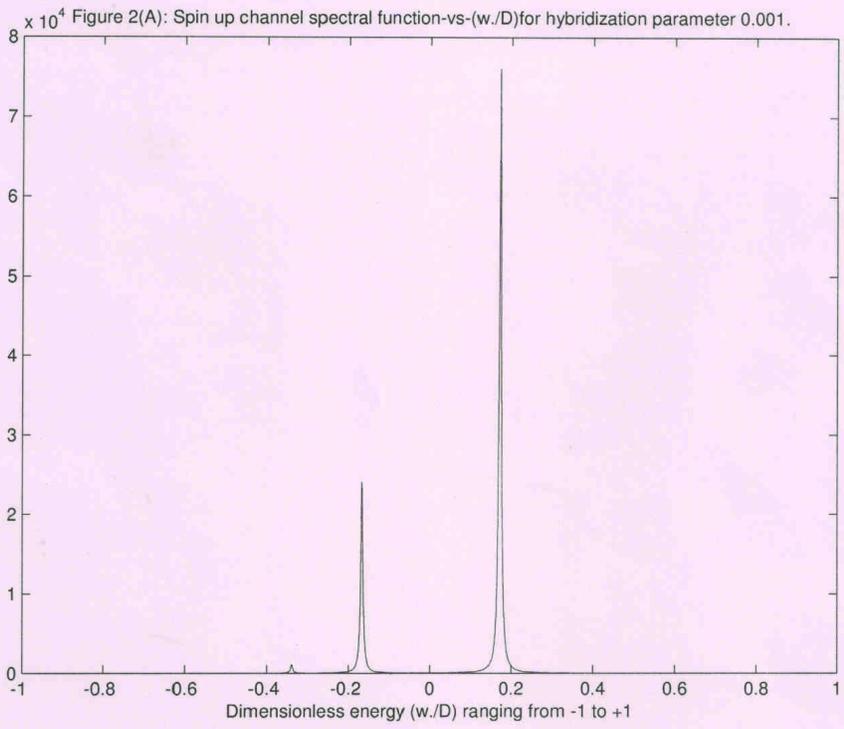

Figure 2(A): Spin up channel spectral function-vs-(w./D) for hybridization parameter 0.001.
Dimensionless energy (w./D) ranging from -1 to +1

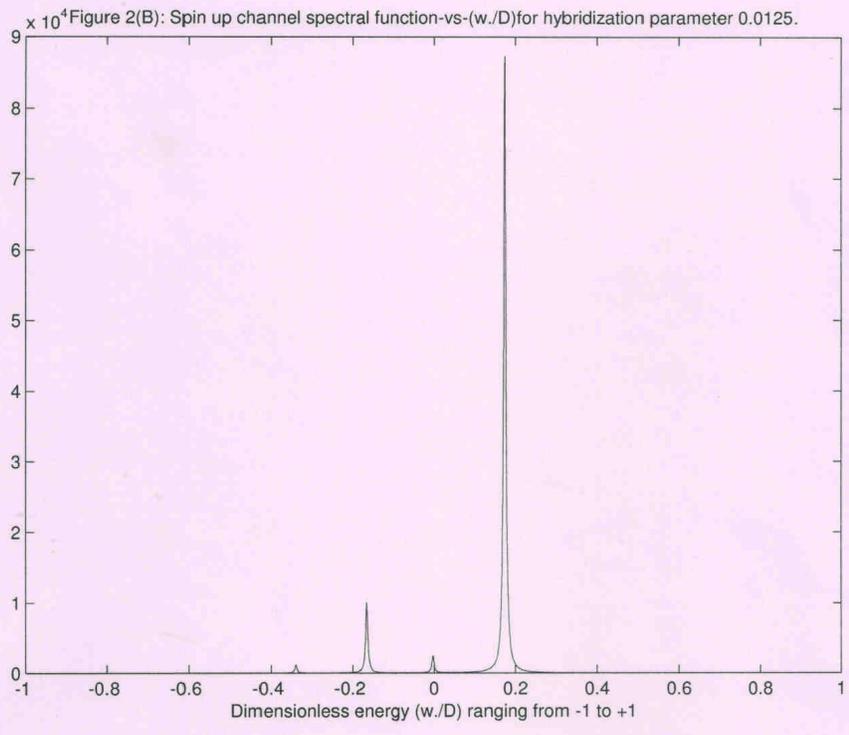

Figure 2(B): Spin up channel spectral function-vs-(w./D) for hybridization parameter 0.0125.
Dimensionless energy (w./D) ranging from -1 to +1

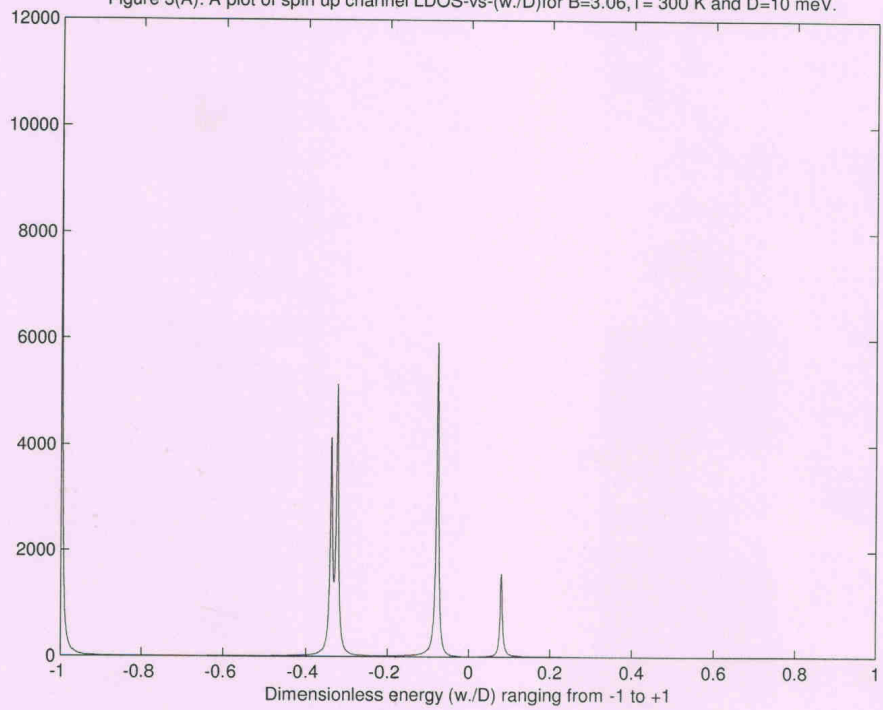

Figure 3(A): A plot of spin up channel LDOS-vs-(w./D) for B=3.06, T= 300 K and D=10 meV.

Dimensionless energy (w./D) ranging from -1 to +1

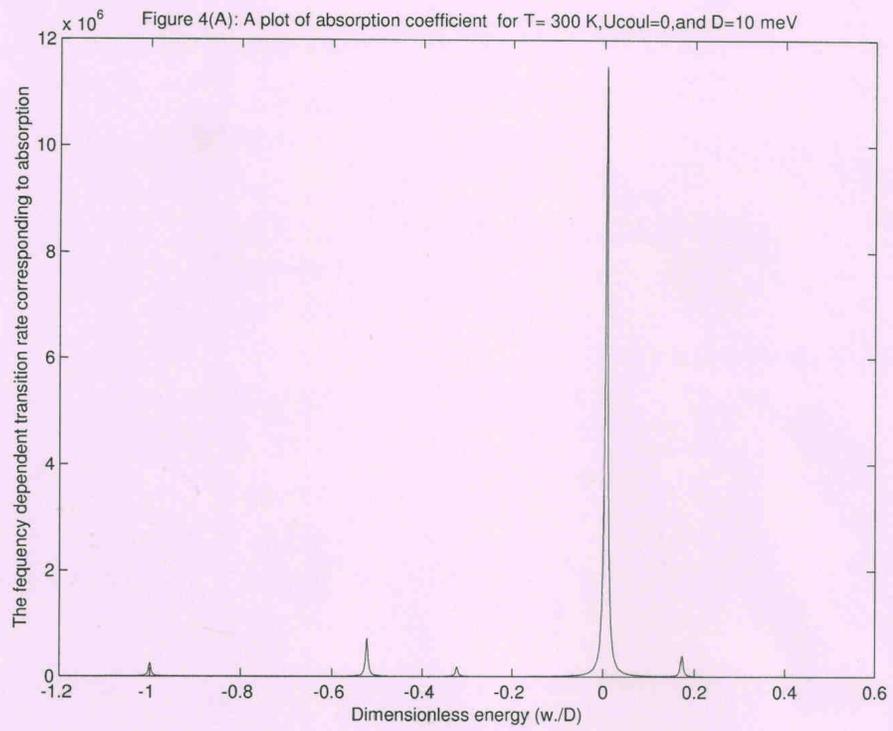

Figure 4(A): A plot of absorption coefficient for T= 300 K, Ucoul=0, and D=10 meV

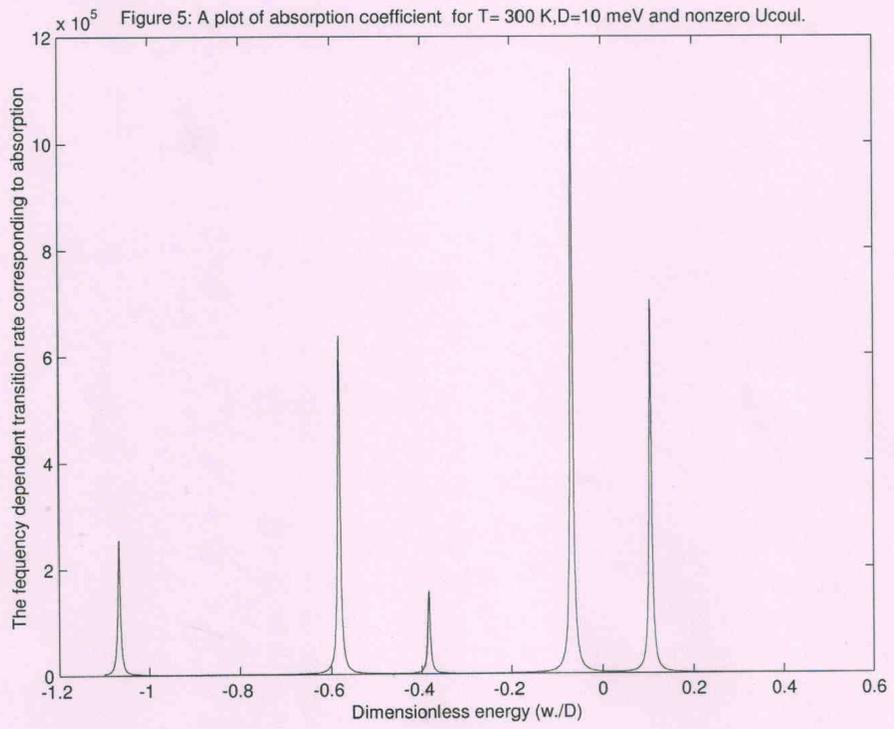

Figure 5: A plot of absorption coefficient for T= 300 K, D=10 meV and nonzero Ucoul.